\documentclass[prl,12pt]{revtex4}

\usepackage{amsmath}
\usepackage{amssymb}

\usepackage{graphicx}

\begin{document}

\title{Magnetic tunnel junctions with impurities}

\author{F. Kanjouri$^{1,2} $, N. Ryzhanova$^{1,3}$, B.  Dieny$^{3}$, N. Strelkov$^{1,3}$, A. Vedyayev$^{1,3}$}
\address{
$^1$Department of Physics, Moscow Lomonosov University, Moscow 119899, Russia\\
$^2$ Department of Physics, Yazd University, Yazd, Iran\\
$^3$SPINTEC, Unit? de Recherche Associ?e 2512 CEA/CNRS,
CEA/Grenoble, D?partement de Recherche Fondamentale sur la Mati?re
Condens?e, 38054 Grenoble Cedex, France}

\begin{abstract}
\textbf{Abstract:} The influence on the I-V characteristics and
tunnel magnetoresistance (TMR), of impurities embedded into the
insulating barrier ($\mathbb{I}$)separating the two ferromagnetic
electrodes ($\mathbb{F}$) of a magnetic tunnel junction, was
theoretically investigated. When the energy of the electron's
bound state at the impurity site is close to the Fermi energy, it
is shown that the current and TMR are strongly enhanced in the
vicinity of the impurity. If the position of the impurity inside
the barrier is asymmetric, e.g. closer to one of the interfaces
${\mathbb F }$/${\mathbb I }$, the I-V characteristic exhibits a
quasidiode behavior. The case of a single impurity and of a random
distribution of impurities within a plane were both studied.
\end{abstract}

\maketitle

Magnetic tunnel junctions (MTJ), consist of two metallic
ferromagnetic electrodes separated by an insulating barrier. They
typically exhibit tunnel magnetoresistance (TMR) of the order of
50\% associated with a change in the relative orientation of the
magnetization in the two ferromagnetic electrodes.  They attract a
lot of attention~\cite{Moodera-Kinder_1,Parkin,Moodera-Mathon_1}
especially due to their applications in several spin-electronic
devices especially in non-volatile MRAM (Magnetic Random Access
Memory). In a pioneer paper~\cite{Slonczewski}, a theory of TMR
for ideal MTJ (without defect) was developed. Later on, it has
been shown~\cite{Tsymbal-Pettifor_1,Vedy-Bagrets_1} that the
presence of different types of defects within the barrier can
dramatically affect the I-V characteristics and TMR amplitude. In
these papers, the current, averaged over the cross-section of the
system, was calculated. However, it is also interesting to
investigate the local current density and TMR in the vicinity of
the impurity. From an experimental point of view, this is
achievable by using conductive Atomic Force Microscopy approach
 as realized for instance in the following reference ~\cite{Da Costa-Henry} where the authors mapped out the spatial variations of the I(V) characteristics
through a tunnel barrier. From a theroretical point of view, a
theory of local impurity assisted tunnelling in MTJ was recently
developed~\cite{Tsymbal-Pettifor_2}. Tight binding model and Kubo
formalism were used to calculate the spin-dependent tunnel current
through the MTJ. In this earlier paper, the I-V characteristics
were not investigated in detail. Furthermore, the dependence of
spin-dependent current on the position of cross section plane
relative to the position of impurity was not calculated.

In the present paper, we report on a theoretical study of the
spatial distribution of spin-dependent current across the plane of
a magnetic tunnel junction. The local I-V charactertistics as well
as the local TMR amplitude are calculated for a single impurity
and for a random planar distribution of impurities inside the
barrier. In this theory, We adopted the free electron model with
exchange splitting for the ferromagnetic electrodes and used the
nonequilibrium Keldysh technique~\cite{Keldysh_eng} to calculate
the transport properties which are nonlinear functions of the
applied voltage.

The MTJ is described as a three layers system, consisting of two
thick ferromagnetic electrodes ${\mathbb F }$ separated by an
insulating layer, ${\mathbb I }$. Inside the barrier, a single
nonmagnetic impurity with attracting potential is located at a
given distance from the ${\mathbb F }$/${\mathbb I }$ interface.
The two cases of parallel and antiparallel orientations of the
${\mathbb F }$-layers magnetization were investigated.

The ${\mathbb F }$-electrodes are connected to the reservoirs with
chemical potentials $\mu_{1}$ and $\mu_{2}$ so that
$\mu_{2}-\mu_{1}=eV$, where $V$ is the applied voltage.

To calculate the current through the system, the Keldysh  Green
function $G^{-+}$ and advanced and retarded Green functions
$G^{A}$ and $G^{R}$ must be calculated. By solving the Dyson
equation, we found that
\begin{multline}
G^{-+}(\textbf{r},\textbf{r}')=G_{0}^{-+}
(\textbf{r},\textbf{r}')+
\frac{G_{0}^{R}(\textbf{r},\textbf{r}_{0})
WG_{0}^{-+}(\textbf{r}_{0},\textbf{r}')}
{1-WG_{0}^{R}(\textbf{r}_{0},\textbf{r}_{0})}
+\frac{G_{0}^{-+}(\textbf{r},\textbf{r}_{0})
WG_{0}^{A}(\textbf{r}_{0},\textbf{r}')}
{1-WG_{0}^{A}(\textbf{r}_{0},\textbf{r}_{0})}
\\+\frac{G_{0}^{R}(\textbf{r},\textbf{r}_{0})
WG_{0}^{-+}(\textbf{r}_{0},\textbf{r}_{0})
WG_{0}^{A}(\textbf{r}_{0},\textbf{r}')}
{\left(1-WG_{0}^{R}(\textbf{r}_{0},\textbf{r}_{0})\right)
\left(1-WG_{0}^{A}(\textbf{r}_{0},\textbf{r}_{0})\right)}\label{Dyson}
\end{multline}
where $G_{0}^{-+} (\textbf{r},\textbf{r}'),G_{0}^{A}
(\textbf{r},\textbf{r}')$ and $G_{0}^{R} (\textbf{r},\textbf{r}')$
are the Green's functions for the system in the absence of the
impurity and the potential of the impurity $V$ was represented as
a $\delta$-function:
$V(\textbf{r})=Wa_{0}^3\delta(z-z_{0})\delta$(\mbox{\boldmath
$\rho$}-\mbox{\boldmath $\rho_{0}$}),
$\textbf{r}_{0}=(\mbox{\boldmath $\rho_{0}$},z_{0})$ is the
position of the impurity, $a_{0}$ is its effective radius, $W$ is
its amplitude. The explicit expressions for $G^A, G^R, G^{-+}$
have the following forms:

\begin{eqnarray}
\begin{array}{l}
G_{0}^{R}(\textbf{r},\textbf{r}')=\int
d^2\kappa\frac{(-1)\mathrm{e}^{-i\mbox{\boldmath$\kappa$}(\mbox{\boldmath$\rho$}-\mbox{\boldmath$\rho$}')}}
{2\sqrt{q(z)q(z')}den}\left\{
  E(z_{2},z)\left[q(z_{2})+ik_{2}\right]+E^{-1}(z_{2},z)\left[q(z_{2})-ik_{2}\right]
  \right\}\\ \times\left\{
  E(z',z_{1})\left[q(z_{1})+ik_{1}\right]+E^{-1}(z',z_{1})\left[q(z_{1})-ik_{1}\right]
  \right\},\label{G0R}
 \end{array}
\end{eqnarray}

\begin{eqnarray}
\begin{array}{l}
G_{0}^{A}(\textbf{r},\textbf{r}')=\int d^2\kappa\frac{(-1)
  \mathrm{e}^{i\mbox{\boldmath$\kappa$}(\mbox{\boldmath$\rho$}-\mbox{\boldmath$\rho$}')}}{2\sqrt{q(z)q(z')}den^*}\left\{
  E(z_{2},z)\left[q(z_{2})-ik_{2}\right]+E^{-1}(z_{2},z)\left[q(z_{2})+ik_{2}\right]
  \right\}\\ \times\left\{
  E(z',z_{1})\left[q(z_{1})-ik_{1}\right]+E^{-1}(z',z_{1})\left[q(z_{1})+ik_{1}\right]
  \right\},\label{G0A}
 \end{array}
\end{eqnarray}

\begin{eqnarray}
\begin{array}{l}
G_{0}^{-+}(\textbf{r},\textbf{r}')=\int
d^2\kappa\frac{i4k_{1}q(z_{1})n_{L}\mathrm{e}^{-i\mbox{\boldmath$\kappa$}(\mbox{\boldmath$\rho$}-\mbox{\boldmath$\rho$}')}}
{\sqrt{q(z)q(z')}|den|^2}\left\{
  E(z',z_{2})\left[q(z_{2})+ik_{2}\right]+E^{-1}(z',z_{2})\left[q(z_{2})-ik_{2}\right]
  \right\}\\ \times\left\{
  E(z,z_{2})\left[q(z_{2})-ik_{2}\right]+E^{-1}(z,z_{2})\left[q(z_{2})+ik_{2}\right]
  \right\}\\
  \\+\int
d^2\kappa\frac{i4k_{2}q(z_{2})n_{R}\mathrm{e}^{-i\mbox{\boldmath$\kappa$}
(\mbox{\boldmath$\rho$}-\mbox{\boldmath$\rho$}')}}{\sqrt{q(z)q(z')}|den|^2}\left\{
  E(z_{1},z')\left[q(z_{1})+ik_{1}\right]+E^{-1}(z_{1},z')\left[q(z_{1})-ik_{1}\right]
  \right\}\\ \times\left\{
  E(z_{1},z)\left[q(z_{1})-ik_{1}\right]+E^{-1}(z_{1},z)\left[q(z_{1})+ik_{1}\right]
  \right\},\label{G0-+}
 \end{array}
\end{eqnarray}

where
\begin{eqnarray*}
\begin{array}{l}
q(z)=\sqrt{q_{0}^{2}+\kappa^{2}-\frac{2m}{\hbar^2}\frac{(z-z_{1})}{(z_{2}-z_{1})}eV},
\\\\k_{1}=\sqrt{\frac{2m}{\hbar^2}(\varepsilon-\Delta_{1})-\kappa^{2}},
\\\\k_{2}=\sqrt{\frac{2m}{\hbar^2}(\varepsilon-\Delta_{2}+eV)-\kappa^{2}},
\\\\den=\left\{
{E(z_{1},z_{2})\left[q(z_{2})-ik_{2}\right]\left[q(z_{1})-ik_{1}\right]
-E^{-1}(z_{1},z_{2})\left[q(z_{2})+ik_{2}\right]\left[q(z_{1})+ik_{1}\right]}\right\},
\\ \\E(z_{1},z_{2})\equiv \mathrm{e}^{\int_{z_{1}}^{z_{2}}
q(\tau)d\tau},
\end{array}
\end{eqnarray*}
\mbox{\boldmath$\kappa$} is the electron momentum perpendicular to
the plane of structure, $\varepsilon$ is the energy, $z_{1}$ and
$z_{2}$ are the positions of the ${\mathbb F }$/${\mathbb I }$
interfaces, $\Delta_{1}$ and $\Delta_{1}$ denote   the positions
of the bottom of the energy band for spin up and down subbands.

\noindent  $n_L=f^0(\varepsilon)$ and $n_R=f^0(\varepsilon+eV)$
are Fermi distribution functions in the left and right reservoirs
and $\frac{\hbar^2q_{0}^2}{2m}$ height of potential barrier above
Fermi level.

\begin{figure}[ht]
\includegraphics*[width=0.57\textwidth]{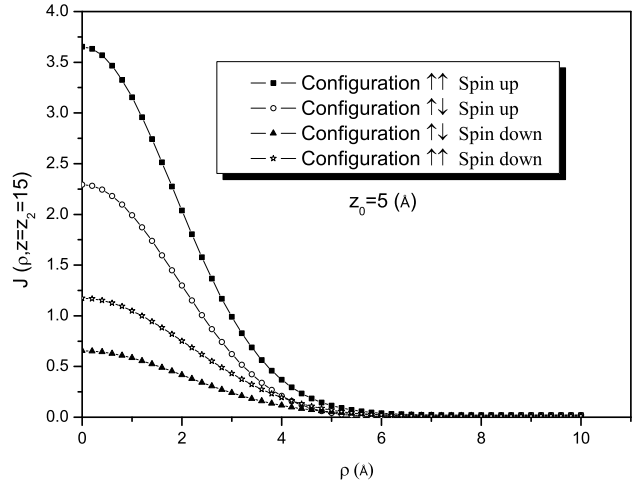}
\caption{Dependence of the current for different spin channels and
P and AP configuration on the distance from the impurity in the
plane of the structure at z=15 \AA. $k_{F}^\uparrow=1.1
~{\mbox\AA}^{-1}$, $k_{F}^\downarrow=0.6 ~{\mbox\AA}^{-1}$,
$q_{0}=1.0 ~{\mbox\AA}^{-1}$ }. \label{fig:fig1}
\end{figure}
\begin{figure}[h]
\includegraphics*[width=0.57\textwidth]{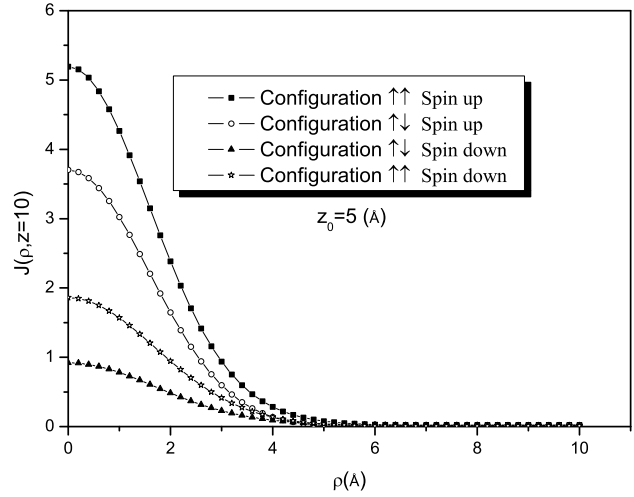}
\caption{The same dependence at $z=10$ \AA .} \label{fig:fig2}
\end{figure}

\begin{figure}[h]
\includegraphics*[width=0.57\textwidth]{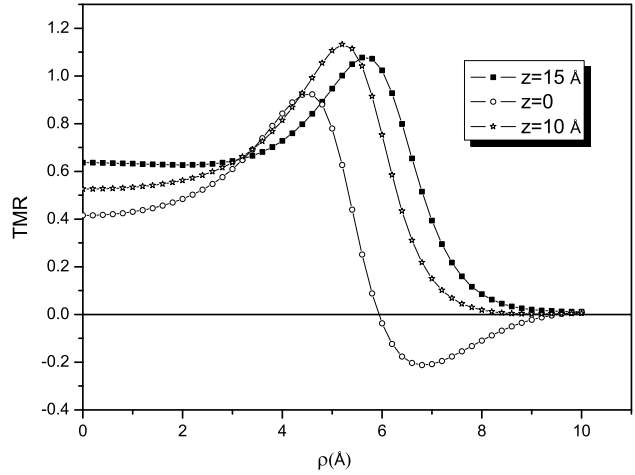}
\caption{Dependence of TMR on the distance from the impurity in
the plane of the structure at different $z$. For parameters see
Fig.\ref{fig:fig1}} \label{fig:fig3}
\end{figure}

\begin{figure}[h]
\includegraphics*[width=0.5\textwidth,angle=-90]{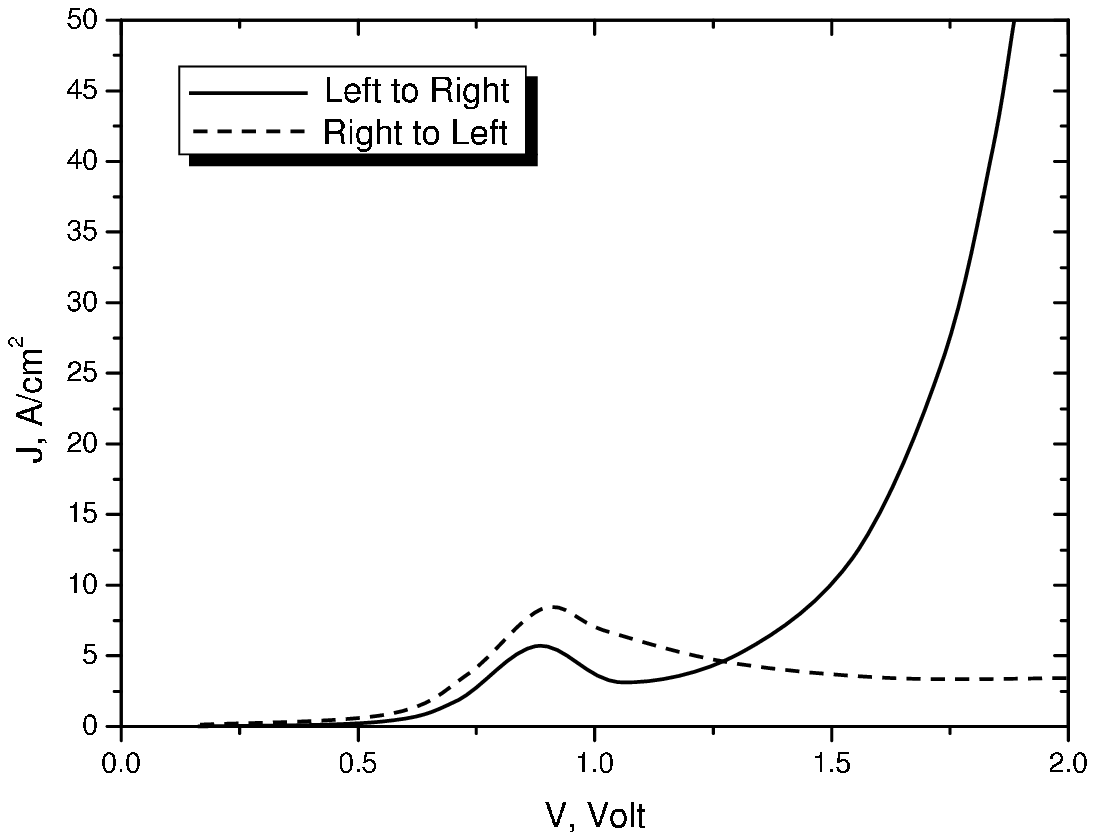}
\caption{ Local I-V curve at $\rho=\rho_{0}$ and $z=15$ \AA ~for
the case of single impurity.} \label{fig:fig4}
\end{figure}
\begin{figure}[ht]
\includegraphics*[width=0.5\textwidth,angle=-90]{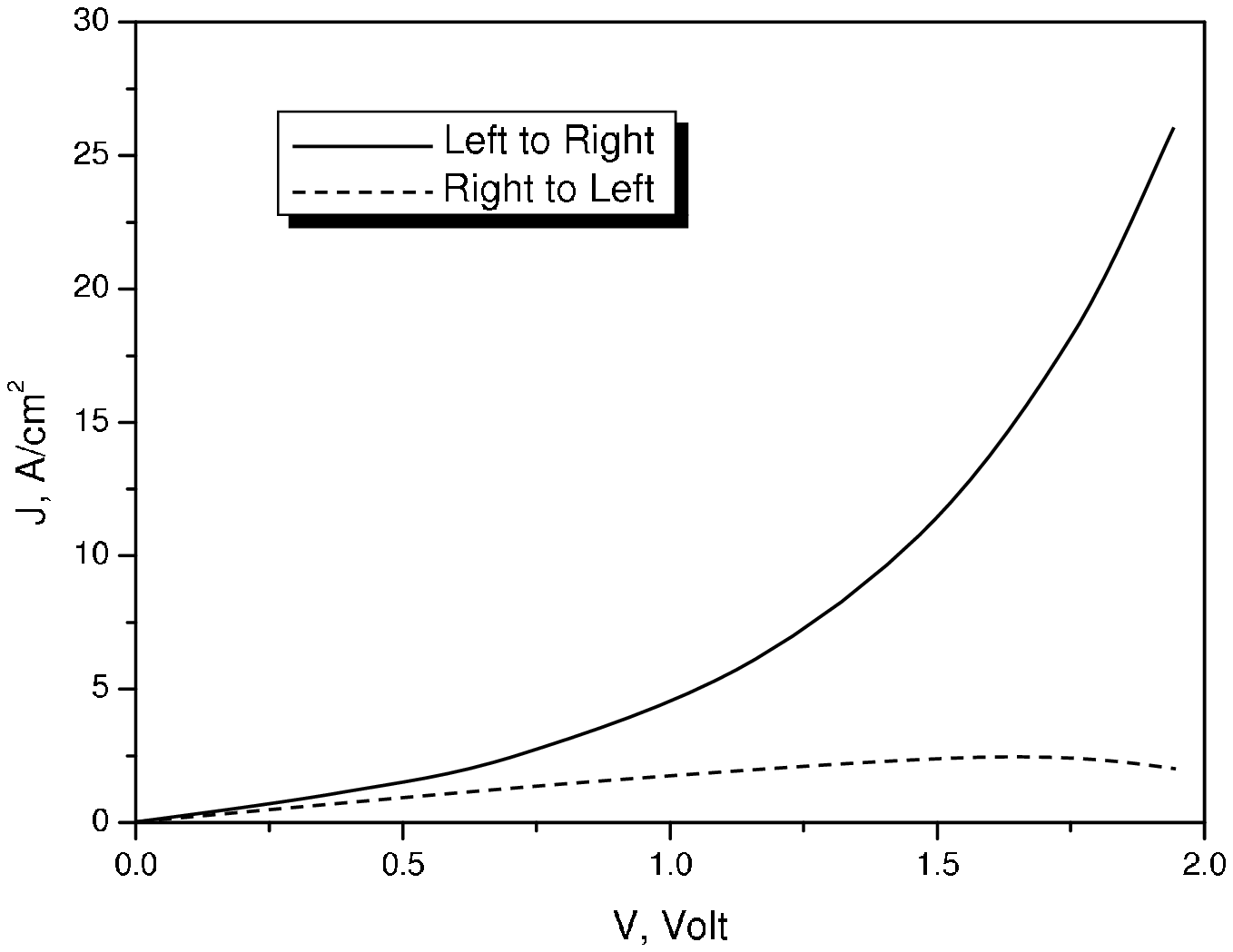}
\caption{I-V curve in the case of the layer of impurities at
$z_{0}=3$\AA ~and $x=0.5$.} \label{fig:fig5}
\end{figure}

In (\ref{Dyson}),(\ref{G0R}),(\ref{G0A}) and (\ref{G0-+})
\mbox{\boldmath$\rho$} and $z$ are in the plane and perpendicular
to the plane coordinates, and we consider that $z$ and $z_{0}$ are
situated within the barrier. We have to take into account that all
Green functions are matrices in spin space. $k_{1F}^\uparrow$,
$k_{1F}^\downarrow$, $k_{2F}^\uparrow$, $k_{2F}^\downarrow$ are
Fermi wave vectors of electron with spin$\uparrow(\downarrow)$ in
the left and right $\mathbb{F}$-electrodes. The current was
calculated, using the following expression:

\begin{equation}
j_{z}(\rho,z)=\frac{e\hbar}{2m}\int
d\varepsilon\left(\frac{\partial G^{-+}(z,\rho;z',\rho)}{\partial
z'}-\frac{\partial G^{-+}(z,\rho;z',\rho)}{\partial
z}\right)_{z=z'}\label{Jz}
\end{equation}

In Fig.\ref {fig:fig1} and Fig.\ref {fig:fig2},  the dependencies
of the currents in different channels (up and down spin) on
coordinate $\rho-\rho_{0}$ at one interface
$\mathbb{I}$/$\mathbb{F}$ ($z_{2}=15$\AA) (another interface is at
$z_{1}=0$) and inside the barrier at $z=10$ are shown. In this
calculation, the impurity is assumed to be positioned at
$\rho_{0}=0$ and $z_{0}=5$\AA.

In the vicinity of the impurity, a hot spot of radius
approximately equal to 6\AA ~may be observed. The current density
in the center of the hot spot exceeds the value of the background
current by several orders of magnitude. In Fig.\ref {fig:fig3},
the TMR dependence on the distance from the impurity at different
$z$ is shown. It is interesting that the value of TMR in the
vicinity of the impurity exceeds its background value (TMR for the
ideal structure is equal $0.013$) by more than an order of
magnitude. Furthermore, in some cases, regions of $\rho-\rho_{0}$
exist in which the TMR becomes negative.

Fig.\ref {fig:fig4} shows the I-V characteristics for positive and
negative applied voltage. These curves are quite asymmetric with
respect to the sign of the voltage. This asymmetry is related to
the asymmetric positioning of the impurity inside of the barrier.
It is particularly pronounced if the potential of the impurity is
chosen so that the bound (resonance) state of electrons with spin
up is located near the Fermi energy for the positive applied
voltage $=1.2 ~V$, and if this bound state lies below the Fermi
energy for negative voltage. This diode behavior was demonstrated
so far in the case of a single impurity. We next investigate the
case of a finite concentration of impurities.

In this case, we consider the same magnetic tunnel barrier
structure with a monolayer of impurities of finite atomic
concentration $x$, situated closer to one of the ${\mathbb F
}$/${\mathbb I }$ interfaces. To solve the problem, as a first
step, we have to find the coherent potential and effective Keldysh
Green function $G_{\mathrm{eff}}^{-+}$. By solving the Dyson
equation in the Keldysh space, the following expression was
obtained for $G_{\uparrow\uparrow}^{-+AP}$:
\begin{multline}
G^{-+}\left({z,z'}\right)=G_0^{-+}\left({z,z'}\right)+
\frac{{G_0^{-+}\left({z,z_0 }\right)\Sigma^A G_0^A \left({z_0
,z'}\right)}} {{1-G_0^A\left({z_0 ,z_0 }\right)\Sigma^A}}
+\frac{{G_0^R \left({z,z_0}\right)\Sigma^R G_0^{-+} \left({z_0
,z'}\right)}}
{{1-G_0^R \left( {z_0 ,z_0 } \right)\Sigma^R }}-\\
\frac{{G_0^R \left( {z,z_0 } \right)\Sigma^{-+} G_0^A \left({z_0
,z'}\right)}} {{\left({1-G_0^A\left({z_0
,z_0}\right)\Sigma^A}\right)\left({1-G_0^R\left({z_0,z_0}\right)\Sigma^R}\right)}}+
\frac{{G_0^R \left({z,z_0}\right)\Sigma^R
G_0^{-+}\left({z_0,z_0}\right)\Sigma^A G_0^A
\left({z_0,z'}\right)}}
{{\left({1-G_0^A\left({z_0,z_0}\right)\Sigma^A}\right)\left({1-G_0^R\left({z_0,z_0}\right)\Sigma^R}\right)}}\label{G-+}
\end{multline}
where $\Sigma^{R(A)}$ are the coherent potentials (C.P.) for the
retarded and advanced Green functions, which have to be found from
the C.P.A equation:
\begin{equation}
\bar t=(1-x)
\frac{(\varepsilon^A-\Sigma)}{1-(\varepsilon^A-\Sigma)G_{\mathrm{eff}}(z_{0},\rho_{0};z_{0},\rho_{0})}
+(x)\frac{(\varepsilon^B-\Sigma)}{1-(\varepsilon^B-\Sigma)G_{\mathrm{eff}}(z_{0},\rho_{0};z_{0},\rho_{0})}
=0\label{t}
\end{equation}
where $\varepsilon^A$ and $\varepsilon^B$ are the onsite energies
of the host ($Al_2O_3$) and the impurity ($Al$) and
$\Sigma^{-+}=\frac{i}{2}(n_R+n_L)(\Sigma^R-\Sigma^A)$.

Now to calculate the I-V characteristic, we can use the previously
found expression for $G_{\alpha \alpha}^{-+P(AP)}$ and substitute
it into the expression~(\ref{Jz}).

In Fig.\ref{fig:fig5}, the I-V characteristic in the AP
configuration is shown. An asymmetry of the curve on the sign of
the applied voltage is clearly visible.

Such a structure may be prepared for instance by sputtering a thin
layer of $Al$ on the bottom $\mathbb{F}$-electrode, then oxidise
it in Alumina. Thenafter, a second thicker layer of $Al$ is
sputtered on the already formed Alumina barrier but this second
layer is subsequently underoxidized so that a thin layer of the
random alloy $Al_xAl_2O_{3(1-x)}$ remains inside the more or less
ideal insulator $Al_2O_3$ at an assymmetric location within the
barrier.

The work was partly supported by the Russian fund of fundamental
research (grant N 04-02-16688a). AV, NR and NS thank CEA for
financial support during their stay.

\bibliography{impur}

\end{document}